\documentclass[a4paper]{article}

\usepackage{INTERSPEECH2020}

\usepackage{hyperref}
\usepackage{caption}
\usepackage{subcaption}

\usepackage{multirow}
\usepackage[export]{adjustbox}

\title{Preliminary study on using vector quantization latent spaces for TTS/VC systems with consistent performance}
\name{Hieu-Thi Luong$^{1}$, Junichi Yamagishi$^{1}$}
\address{
  $^1$National Institute of Informatics, Tokyo, Japan
  }
\email{\{luonghieuthi,jyamagis\}@nii.ac.jp}
\begin{document}

\maketitle
\begin{abstract}
Generally speaking, the main objective when training a neural speech synthesis system is to synthesize natural and expressive speech from the output layer of the neural network without much attention given to the hidden layers. However, by learning useful latent representation, the system can be used for many more practical scenarios.
In this paper, we investigate the use of quantized vectors to model the latent linguistic embedding and compare it with the continuous counterpart. 
By enforcing different policies over the latent spaces in the training, we are able to obtain a latent linguistic embedding that takes on different properties while having a similar performance in terms of quality and speaker similarity. Our experiments show that the voice cloning system built with vector quantization has only a small degradation in terms of perceptive evaluations, but has a discrete latent space that is useful for reducing the representation bit-rate, which is desirable for data transferring, or limiting the information leaking, which is important for speaker anonymization and other tasks of that nature.

\end{abstract}

\section{Introduction}

When it comes to text-to-speech (TTS) tasks, the deep learning approach has several advantages over the conventional approaches, such as its simple structure and the ability to scale with large data \cite{ping2018deep,sotelo2017char2wav}. These characteristics are key for pushing the performance of speech synthesis systems and machine learning systems in general.
Recent works have shown that a sequence-to-sequence TTS system \cite{ping2018deep,sun2020generating}  trained with a large transcribed speech corpus can synthesize speech with high naturalness directly from text input instead of going through several sub-systems.
While such systems provide a high performance and a straightforward solution for TTS, many researchers have shifted their focus to more elaborate systems that aim to give some of the control back to human users \cite{luong2017adapting,wang2018style}.
For example, Shen et al. \cite{shen2020non} replaced the attention module with a duration prediction to create a more resilient sequence-to-sequence TTS model and enable the ability to control the spacing of generated speech, while Liu et al. \cite{liu2021controllable} explicitly integrated emphatic code into the model so that users can control the emphasis by changing duration, intonation, and energy.

For voice conversion (VC), recent deep learning based systems are formulated around the ability to disentangle speaker and linguistic information of a neural network by using an information bottleneck structure to force the model to learn useful representations \cite{hsu2016voice,wu2020one,sun2020generating}. This information bottleneck structure can simply be a layer with a few units \cite{qian2019autovc}, a variational autoencoder (VAE) model with its encoder's output regularized to approximate a normal distribution \cite{hsu2017voice,kameoka2019acvae}, or a jointly trained discrete latent space through vector quantization \cite{tjandra2019vqvae,ding2019group,wu2020one}.
In these works, the common hypothesis is that using a certain network structure can help train a representation that takes on information and/or properties that are useful for the task at hand.

Previously, we proposed NAUTILUS \cite{luong2020nautilus}, a versatile voice cloning system, that is a fusion of TTS and VC.
By carefully designing the shared and the exclusive components, NAUTILUS can utilize them to perform elaborate tasks such as cloning unseen voices using untranscribed speech. It also has a consistent performance when switching between TTS and VC.
These properties are the result of a unified and robust linguistic latent space achieved by the joint training and the VAE-like structure. However, one may want to use other methods to shape the latent space for many different purposes.
Specifically, if we can train a discrete latent space \cite{jang2016categorical,van2017neural} instead of a continuous one, it will be useful for many applications. For example, a low bit-rate representation \cite{tjandra2019vqvae,dunbar2019zero} is ideal for a client-server VC system in which the speech encoder is stored in the client device while the speech decoder is not. Alternatively, by using the vector quantization bottleneck, we can limit the information getting through the speech encoder, which is important for tasks such as speaker anonymization \cite{fang2019speaker,tomashenko2020introducing} as it helps reduce the leaking of speaker identity through temporal patterns.
In this work, we investigate the use of vector quantization variational autoencoder (VQVAE) \cite{van2017neural} components to model the linguistic latent space of the NAUTILUS system. In addition to conducting experiments to clarify its effect on subjective evaluations, we discuss how different types of assumption about the latent spaces are useful for different scenarios.
We describe the basics of the voice cloning framework with the vector quantization components in Section \ref{sec:methodology} and discuss our motivation in Section \ref{sec:motive}. Section \ref{sec:experiments} lays out the experiment conditions, and Section \ref{sec:evaluations} presents the subjective evaluation results. We conclude in Section \ref{sec:conclusion} with a brief summary of our findings and mention of future works.

\section{Vector Quantization Latent Space for Voice Cloning}
\label{sec:methodology}

\begin{figure}[t]
    \centering
    \includegraphics[width=0.9\columnwidth]{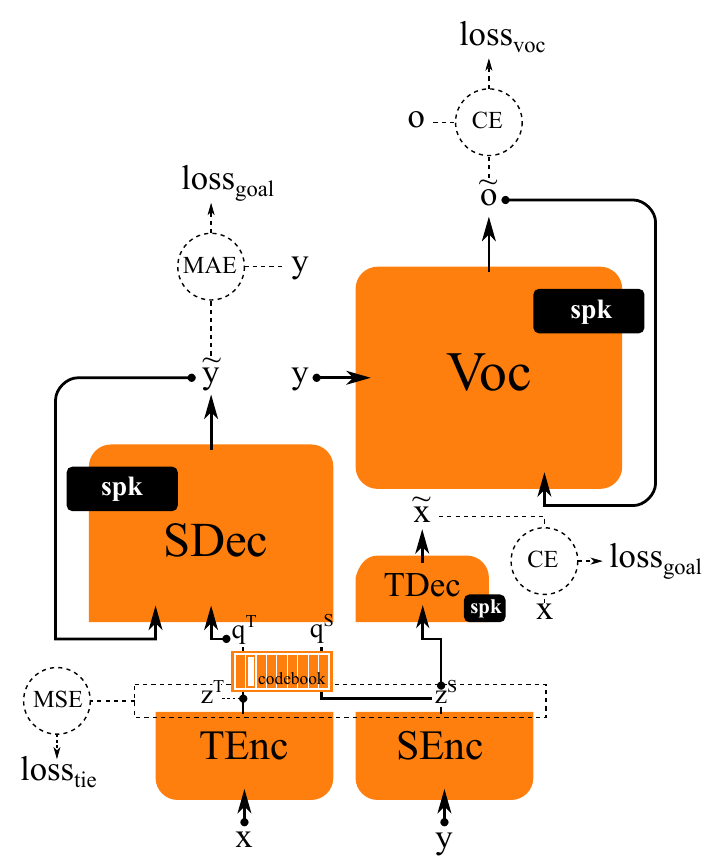}
\vspace{-2mm}
\caption{The modified NAUTILUS-VQ system has a similar structure to the original system, which includes a text encoder ($TEnc$), a speech encoder ($SEnc$), a text decoder ($TDec$), a speech decoder ($SDec$), and a neural vocoder ($Voc$). The jointly trained codebook is the new addition. $\boldsymbol{x}$ is phoneme, $\boldsymbol{y}$ is acoustic, $\boldsymbol{o}$ is waveform, $\boldsymbol{z}$ is continuous latent feature and $\boldsymbol{q}$ is the quantized latent feature. The term $\textrm{loss}_{goal}$ is a placeholder, depending on the encoder/decoder combination --- it can be $\textrm{loss}_{tts}$ (Text-to-speech: $TEnc$\textrightarrow codebook\textrightarrow $SDec$), $\textrm{loss}_{sts}$ (Speech-to-speech, STS: $SEnc$\textrightarrow codebook\textrightarrow $SDec$), $\textrm{loss}_{stt}$ (Speech-to-text, STT: $SEnc$\textrightarrow $TDec$), or $\textrm{loss}_{ttt}$ (Text-to-text: $TEnc$\textrightarrow $TDec$) . The loss functions used in training and adaptation are mean absolute error (MAE), mean squared error (MSE) and cross entropy (CE).}
\label{fig:arc-proposed}
\vspace{-4mm}
\end{figure}

\begin{figure}[t]
    \centering
    \begin{subfigure}[b]{1.0\columnwidth}
         \centering
         \includegraphics[width=0.95\textwidth,right]{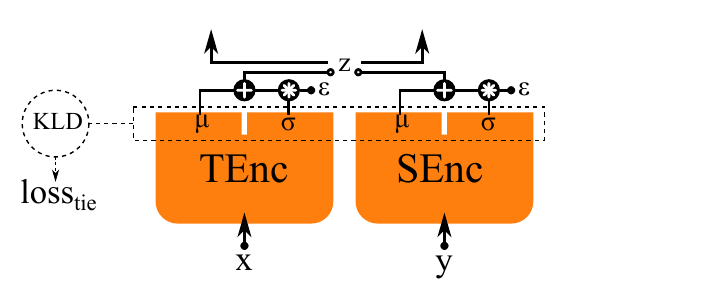}
         \caption{VAE-based encoders}
         \label{fig:encoder-variational}
     \end{subfigure} \\
     \begin{subfigure}[b]{1.0\columnwidth}
         \centering
         \includegraphics[width=0.95\textwidth,right]{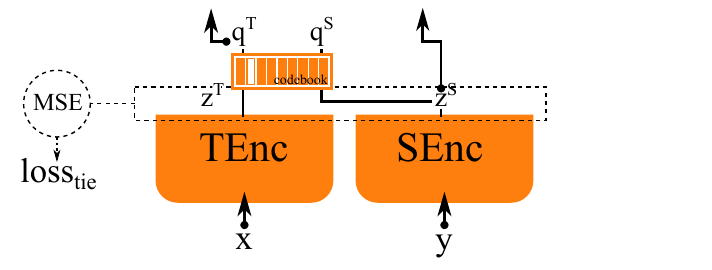}
         \caption{VQVAE-based encoders}
         \label{fig:encoder-quantization}
     \end{subfigure}
\vspace{-5mm}
\caption{The VAE (original) and VQVAE (proposed) setups for the encoders. The VAE-based encoders output mean $\boldsymbol{\mu}$ and standard deviation $\boldsymbol{\sigma}$ and then sample the latent feature $\boldsymbol{z}$ using an $\boldsymbol{\epsilon}$ value drawn from a normal distribution. Kullback-Leibler divergence (KLD) is used as $\textrm{loss}_{tie}$ in this setup. The VQVAE-based encoders output continuous latent feature $\boldsymbol{z}$ that is then quantized into the discrete feature $q$ by using the jointly trained codebook. MSE is used for $\textrm{loss}_{tie}$ in this case.}
\label{fig:arc-encoders}
\vspace{-4mm}
\end{figure}

We adopt the basic concepts of the voice cloning framework proposed in our previous publications \cite{luong2020nautilus} and replace the VAE-based encoders with a VQVAE-based counterpart for this work. Readers may want to refer to the original study \cite{luong2020nautilus} for more context.
Briefly, our voice cloning system is a unified system of TTS and VC, and thanks to this fusion it has the capacity to clone new voices using untranscribed speech. The proposed VQVAE-based system, called NAUTILUS-VQ, is illustrated in Fig. \ref{fig:arc-proposed}. The only difference from the original \cite{luong2020nautilus} is the way the text and speech encoders are set up as shown in Fig. \ref{fig:arc-encoders}. More specifically, the vector quantization bottleneck transforms the continuous latent feature $z$, emitted by the text or speech encoder, into a discrete latent feature $q$ using the jointly trained codebook $\boldsymbol{e}_k, k \in 1...K$, with $q=e_k$ where $k=argmin_j||z-e_j||$. The speech decoder then consumes $q$, instead of $z$, to reconstruct the acoustic feature $y$ that is used to synthesize speech waveform $o$.

\subsection{Train the initial model}

First, we need to jointly train the text/speech encoders/decoders and the codebook in a supervised fashion by using a large-scale transcribed multi-speaker speech corpus and optimizing a designated loss:
\begin{equation}
\begin{aligned}
\label{eq:losstrain}
\textrm{loss}_{train} &= \textrm{loss}^{tts}_{train} + \alpha_{sts} \; \textrm{loss}^{sts}_{train} + \alpha_{stt} \; \textrm{loss}_{stt} \\
                      &\qquad + \beta \; \textrm{loss}_{tie} \;.
\end{aligned}
\end{equation}
Please see the caption of Fig.\ \ref{fig:arc-proposed} for subscripts of each term. The basic structure of the training loss is not much different from the original \cite{luong2020nautilus}, but, due to the vector quantization components, the details are a little more complex. Specifically $\textrm{loss}^{tts}_{train}$ is similar to a typical VQVAE setup \cite{van2017neural}:
\begin{equation}
\begin{aligned}
\label{eq:losstraintts}
\textrm{loss}^{tts}_{train} &= \textrm{loss}_{tts} + \delta_{VQ} \; \textrm{loss}^{T}_{VQ} + \delta_{C} \; \textrm{loss}^{T}_{C} \;.
\end{aligned}
\end{equation}
where $\textrm{loss}_{tts} = || \tilde{y}^T - y ||$ is the reconstruction loss of the acoustic feature, $\textrm{loss}^{T}_{VQ} = || sg(z^T) - q^T ||^2_2$ is the codebook training loss in response to the text input, $x$, from the text encoder, and $\textrm{loss}^{T}_{C} = || z^T - sg(q^T) ||^2_2$ is the text encoder commitment. The operator $sg()$ indicates the stop gradient operation. Similarly, we have the optimization loss for the STS stack that handles the speech input:
\begin{equation}
\begin{aligned}
\label{eq:losstrainsts}
\textrm{loss}^{sts}_{train} &= \textrm{loss}_{sts} + \delta_{VQ} \; \textrm{loss}^{S}_{VQ} + \delta_{C} \; \textrm{loss}^{S}_{C} \;.
\end{aligned}
\end{equation}
Unlike the VAE-based setup \cite{luong2020nautilus} which used Kullback-Leibler (KL) divergence to ``tie'' the encoders' outputs, the VQVAE-based system uses MSE as the latent tying loss:
\begin{equation}
\label{eq:losstie}
    \textrm{loss}_{tie} = ||sg(z^T)-z^S||^2_2 \;.
\end{equation}
Note that we stop the gradient on the text-encoded latent feature, which basically creates an asymmetric tied-layer loss instead of the symmetric KL divergence function as in the original \cite{luong2020nautilus}. A multi-speaker WaveNet vocoder, unchanged from the original, is separately initialized using the same corpus:
\begin{equation}
    \textrm{loss}^\prime_{train} = \textrm{loss}_{voc} \; ,
\end{equation}
The speech decoder, text decoder, and neural vocoder contain speaker dependent (SD) components, which are just one-hot vectors representing speakers in the training set. These SD components will be removed in the voice cloning steps along with the text decoder, which is only included as an auxiliary phone classification regularizer \cite{luong2020nautilus}.

\begin{figure*}[t]

\begin{subfigure}[b]{0.31\textwidth}
         \centering
         \includegraphics[width=\textwidth]{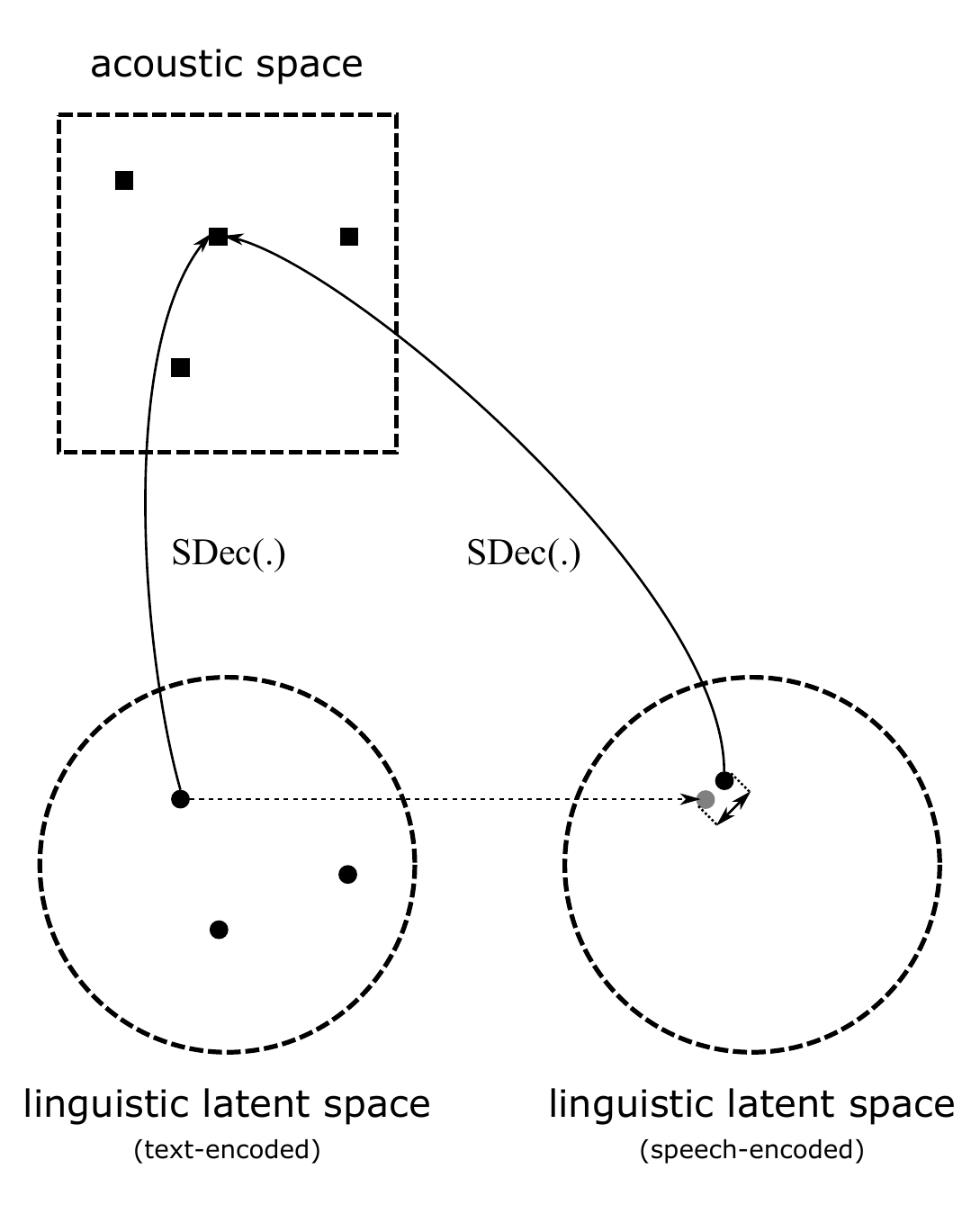}
         \caption{Standard}
         \label{fig:assumption-standard}
     \end{subfigure}
     \hfill
     \begin{subfigure}[b]{0.31\textwidth}
         \centering
         \includegraphics[width=\textwidth]{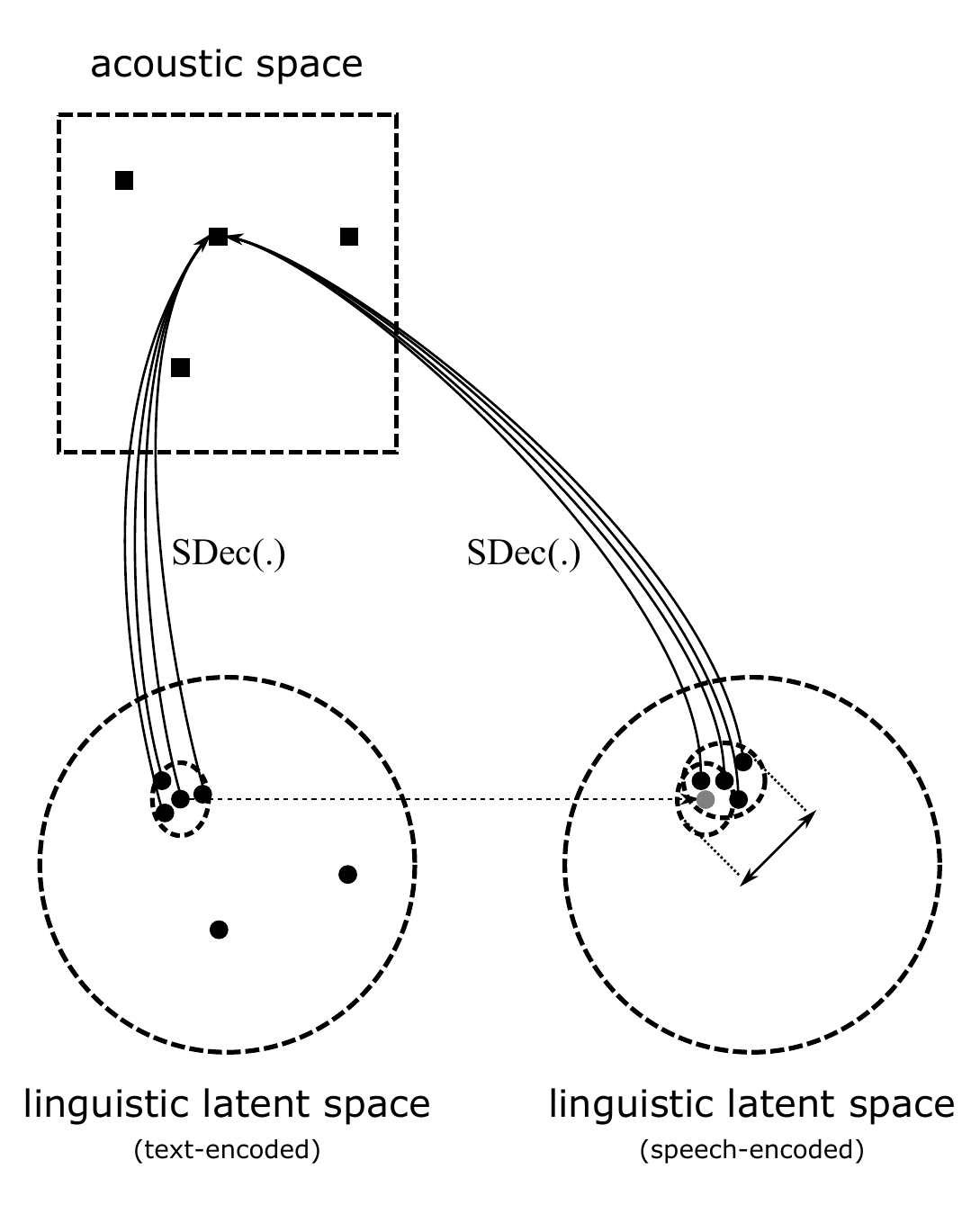}
         \caption{VAE-based}
         \label{fig:assumption-variational}
     \end{subfigure}
     \hfill
     \begin{subfigure}[b]{0.31\textwidth}
         \centering
         \includegraphics[width=\textwidth]{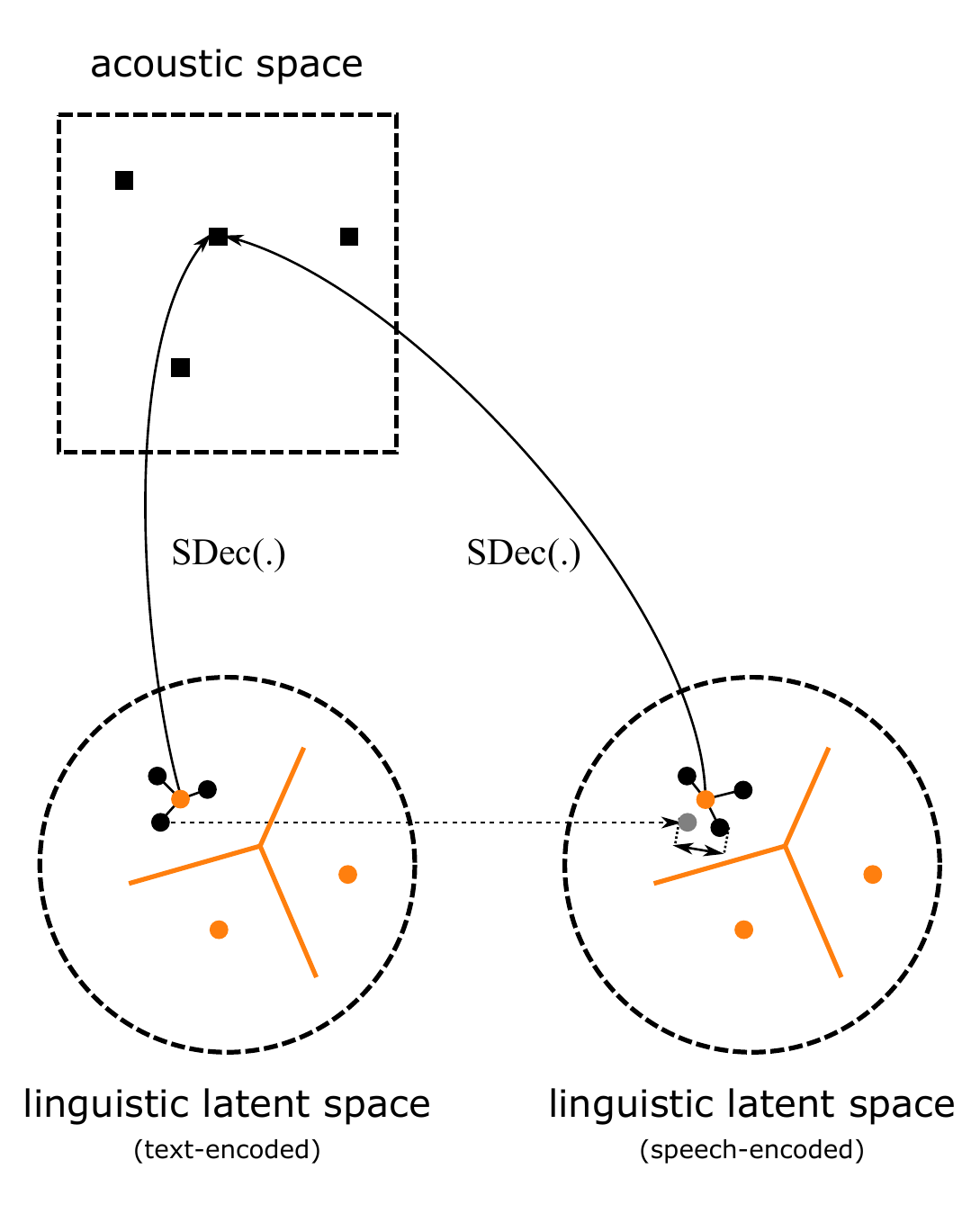}
         \caption{VQVAE-based}
         \label{fig:assumption-vq}
     \end{subfigure}
    \vspace{-2mm}
    \caption{Different ways the linguistic latent spaces are setup and the methods used to enforce the consistency between them.}
    \label{fig:assumption}
  \centering
\vspace{-4mm}
\end{figure*}

\subsection{Clone the target voice}
Given the untranscribed speech of an unseen speaker, we adapt the initial model to generate speech with the voice of the new target, using either the TTS or VC interface.
\subsubsection{Step 1 - Adaptation}
After removing all SD components, we use the STS stack ($SEnc$\textrightarrow codebook\textrightarrow $SDec$) to fine-tune the speech decoder while keep other modules immutable:
\begin{equation}
    \textrm{loss}_{adapt} = \textrm{loss}_{sts} \; ,
\end{equation}
The neural vocoder is also adapted using the same strategy (removing SD components, fine-tuning the rest):
\begin{equation}
    \textrm{loss}^\prime_{adapt} = \textrm{loss}_{voc} \; .
\end{equation}
As this step is similar to the original framework, reader can refer to Fig.\ 2a in \cite{luong2020nautilus} for details.

\subsubsection{Step 2 - Welding}
The adapted model obtained in step one is already capable of generating speech with the target voice. However, we want to increase the speech decoder and the neural vocoder compatibility by jointly tuning them using the speech to waveform stack ($SEnc$\textrightarrow codebook\textrightarrow $SDec$\textrightarrow $Voc$):
\begin{equation}
    \textrm{loss}_{weld} = \textrm{loss}_{sts} + \gamma \; \textrm{loss}_{voc} \; .
\end{equation}
The $\textrm{loss}_{sts}$ is included to maintain the acoustic space for the autoregressive speech decoder  (see Fig.\ 2b in \cite{luong2020nautilus}). 

\subsubsection{Step 3 - Inference}
After the previous steps, the adapted model can generate speech with the voice of the target. The TTS stack and the neural vocoder form a TTS interface that transforms phoneme sequences into a speech waveform, while the STS stack and the neural vocoder form a VC interface that converts utterances spoken by arbitrary source speakers into speech of the same content but with the voice of the target (Fig.\ 2c in \cite{luong2020nautilus}).

\section{Shaping the linguistic latent spaces}
\label{sec:motive}

Our voice cloning system functions on the assumption of a robust and consistent linguistic latent space. Briefly, we want the latent linguistic embedding (LLE) to contain linguistic information, which is useful for speech reconstruction, but none of the speaker information. 
Moreover, the consistency between the TTS and VC interfaces is dictated by the consistency between the text-encoded and speech-encoded latent spaces. In other words, the proposed system achieves a perfect consistency if its text and speech encoders produce an identical LLE sequence given a sentence input and a spoken utterance of the same content.
Note that, for many practical applications \cite{luong2020latent}, it may not even be desirable to achieve such consistency as it eliminates the ability to synthesize speech content that cannot be represented in written form. However, establishing a straightforward goal about consistency helps to simplify the analysis.

\begin{table}[tb]
    \caption{Japanese target speakers for voice cloning task}
\vspace{-2mm}
    \centering
    \scalebox{1.0}{
    \begin{tabular}{llllrr}
        \hline \hline
        Speaker & Gender & Quantity & Duration \\\hline
        F001 & female & 483 utt. & 45.0 min \\
        F002 & female & 481 utt. & 44.4 min \\
        F003 & female & 484 utt. & 47.4 min \\
        F004 & female & 468 utt. & 40.8 min \\
        F005 & female & 485 utt. & 47.6 min\\ \hline
        \multirow{5}{*}{XL10} & \multirow{5}{*}{female} & 10 utt. & 55 s \\
         & & 125 utt. & 10.9 min \\
         & & 500 utt. & 44.5 min \\
         & & 2000 utt. & 2.9 h \\ \cline{3-4}
         & & 8750 utt. & 12.9 h \\ \hline
    \end{tabular}}
    \label{tab:data}
\vspace{-4mm}
\end{table}

We can simply use standard latent features \cite{huang2021pretraining} as LLEs and assume the text and speech encoders produce identical features when consuming different modalities of the same content, and so we optimize the consistency between them by minimizing the distance between the two latent points \cite{luong2018multimodal} as shown in Fig.\ \ref{fig:assumption-standard}. The choice of distance function is another decision that could affect the performance \cite{karita2019semi}, but in practice most use Euclidean distance for its simplicity \cite{zhang2021transfer}.
The flaws of this assumption are the one-to-many relation of text and speech and the scarcity nature of data, which make it difficult to train a robust and consistent latent space.
To address this problem, we utilized the VAE-based encoders (Fig.\ \ref{fig:assumption-variational}) in our previous works \cite{luong2020nautilus}. This modification provides two key benefits: 1) the speech decoder is trained in a denoising fashion due to the sampling process, which can be interpreted as an artificial data argumentation, and 2) we can use a density-wise instead of a point-wise function to connect the text and speech encoders which helps with the consistency.
However, it has the common drawback of the VAE model \cite{kingma2014semi} which is the average-ness of the generated features \cite{sun2020generating}. Therefore, in this paper, we investigate the VQVAE-based modification that has discrete latent spaces, as shown in Fig. \ref{fig:assumption-vq}. The hypothesis is that the discrete features will allow the speech decoder to learn fine-grain details. Moreover, it has several useful traits as mentioned in previous sections.

\section{Experiments}
\label{sec:experiments}

\subsection{Model and training configurations}

For the experiments, we compared the performances of the original NAUTILUS system \cite{luong2020nautilus} and the new system with vector quantization modification. The network architecture of the original was unchanged from the previous publication, and consists of many layers of causal and non-causal dilated convolution layers. Readers can refer to Fig.\ 4 and Sec.\ IV of \cite{luong2020nautilus} for details. The NAUTILUS-VQ system also adopts this architecture but with several changes to reflect the vector quantization components.
Specifically, the encoders directly output 64-dimensional latent vectors instead of means and standard variances. The 160-code jointly trained codebook was used to transform these continuous features into discrete ones. We chose this size for the codebook because it produces a relatively reliable performance based on several test-runs and relevant publications pertaining to VQVAE \cite{ding2019group,williams2021learning}.
For the hyperparameters, we set $\alpha=0.1$, $\beta=0.25$, $\gamma=0.01$, the same as in \cite{luong2020nautilus}, and $\delta_{C}=1.0$, $\delta_{VQ}=0.25$, as based on relevant works \cite{van2017neural,ding2019group}.

\begin{figure}[t!]
    \begin{subfigure}[b]{0.95\linewidth}
        \centering
        \includegraphics[width=\textwidth]{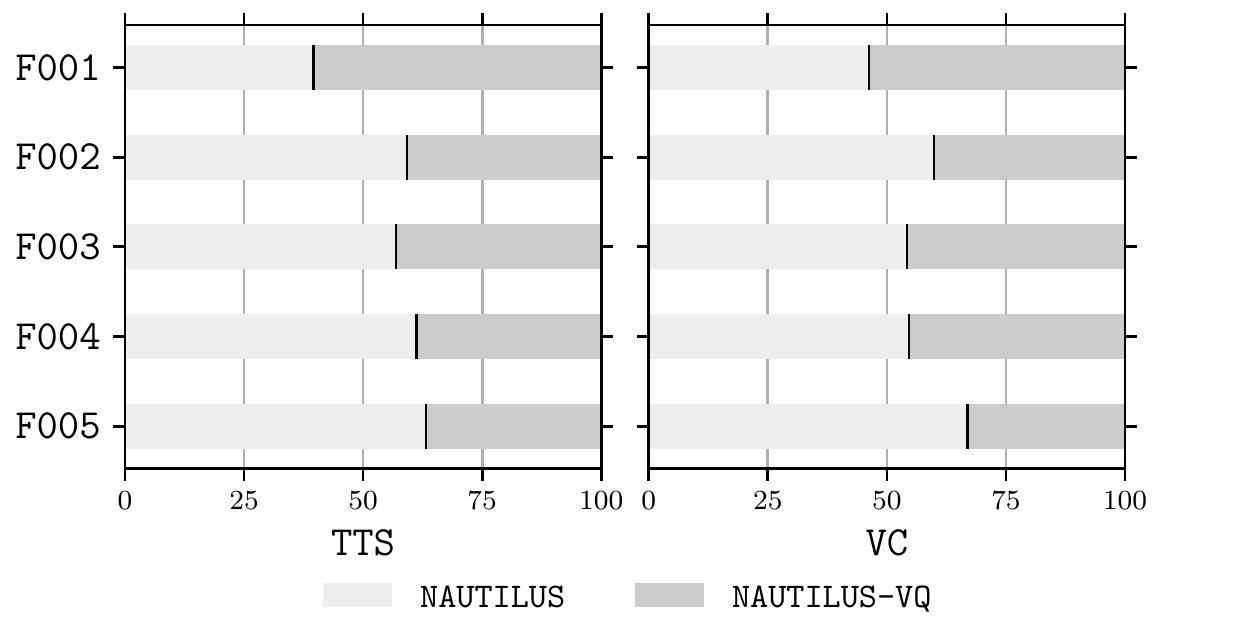}
        \caption{Quality}
        \label{fig:sub-qua}
    \end{subfigure}
    
    \vspace{1mm}
    \begin{subfigure}[b]{0.95\linewidth}
        \centering
        \includegraphics[width=\textwidth]{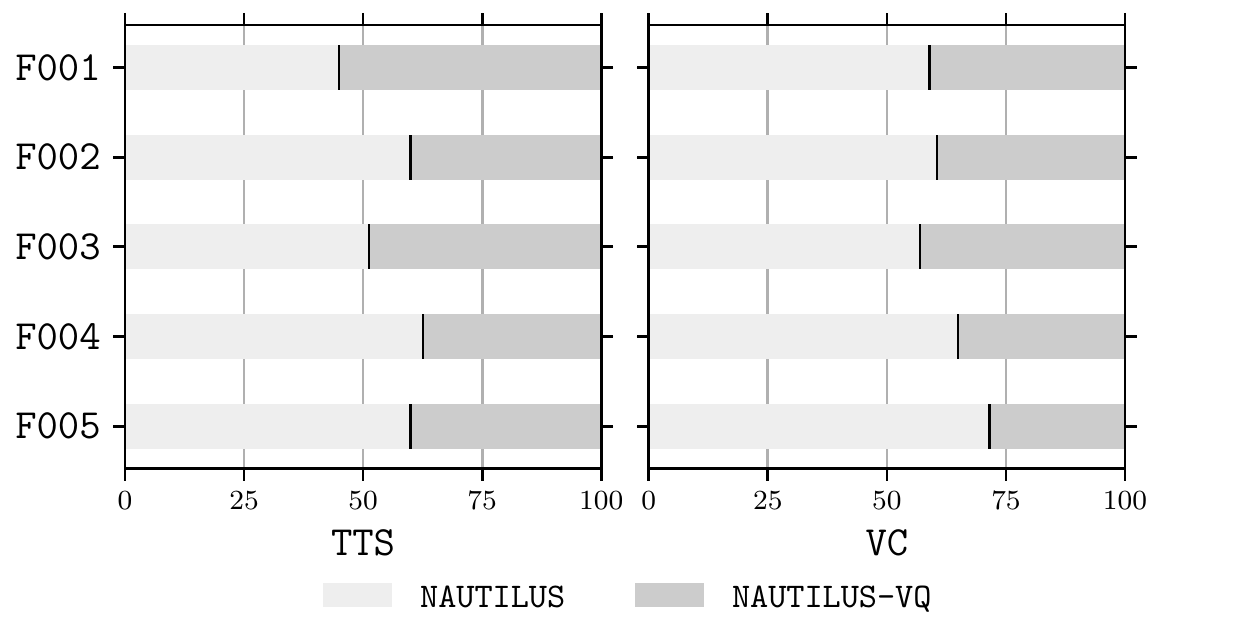}
        \caption{Similarity}
        \label{fig:sub-sim}
    \end{subfigure} 
\vspace{-2mm}
\caption{Subjective evaluations on quality and speaker similarity between NAUTILUS and NAUTILUS-VQ on voice cloning task for TTS and VC.}
\label{fig:sub1}
\vspace{-4mm}
\end{figure}

\begin{figure}[t!]
    \begin{subfigure}[b]{0.95\linewidth}
        \centering
        \includegraphics[width=\linewidth]{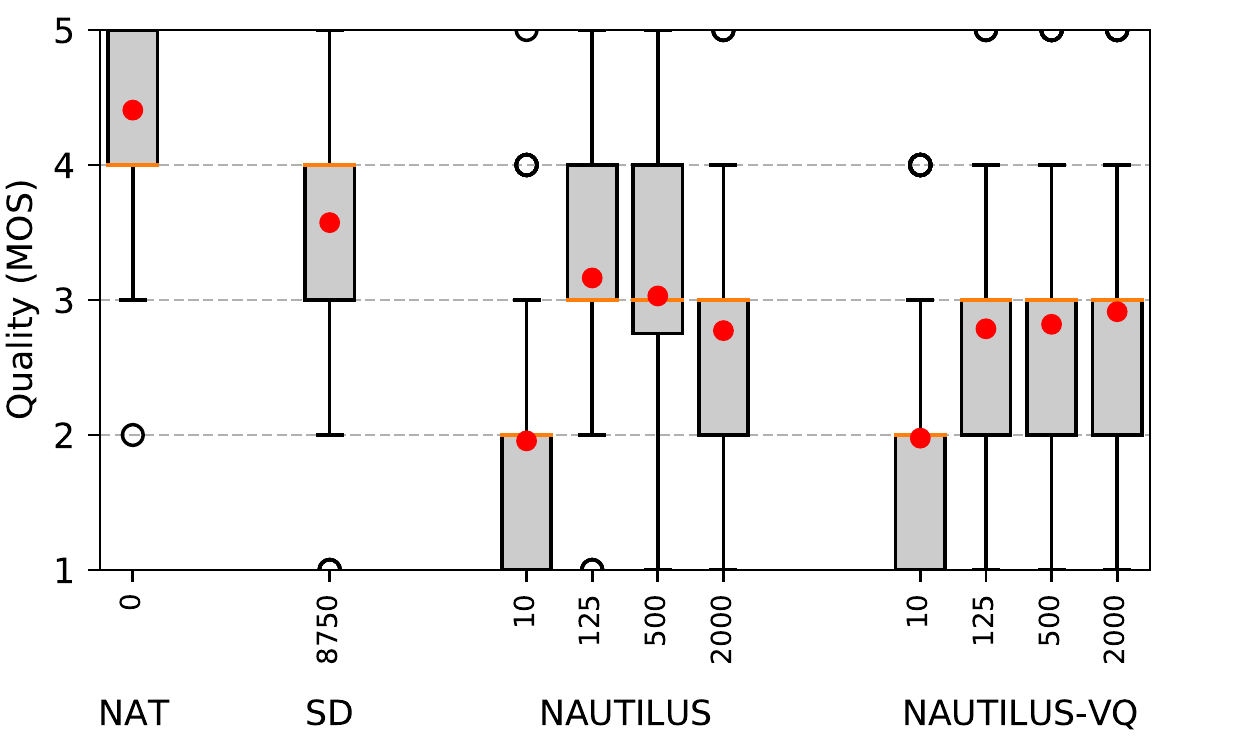}
        \vspace{-5mm}
        \caption{Quality}
    \end{subfigure} \\
    \begin{subfigure}[b]{0.95\linewidth}
        \centering
        \includegraphics[width=\linewidth]{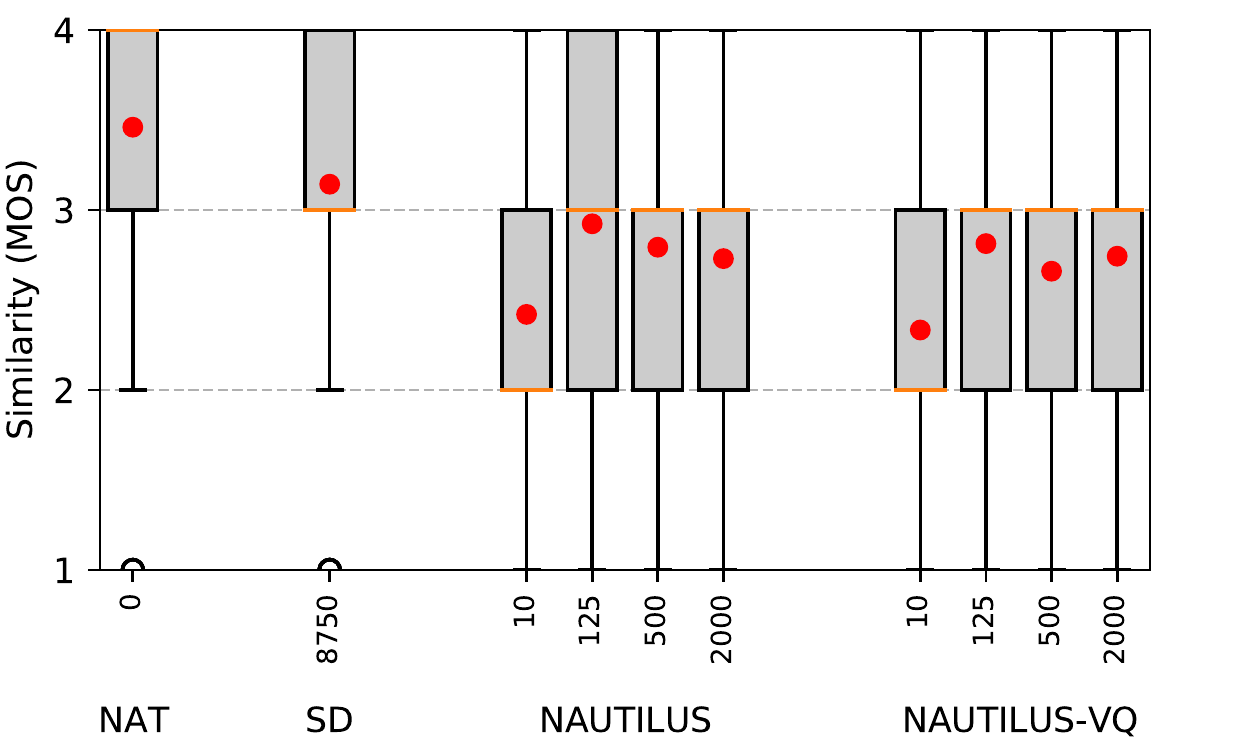}
        \vspace{-5mm}
        \caption{Similarity}
    \end{subfigure}
\vspace{-2mm}
\caption{Subjective evaluations on quality and speaker similarity of the TTS unsupervised speaker adaption using varying amount of untranscribed speech data of speaker XL10.}
\label{fig:sub2}
\vspace{-4mm}
\end{figure}

\begin{figure*}[t]
        \centering
        \includegraphics[width=\linewidth]{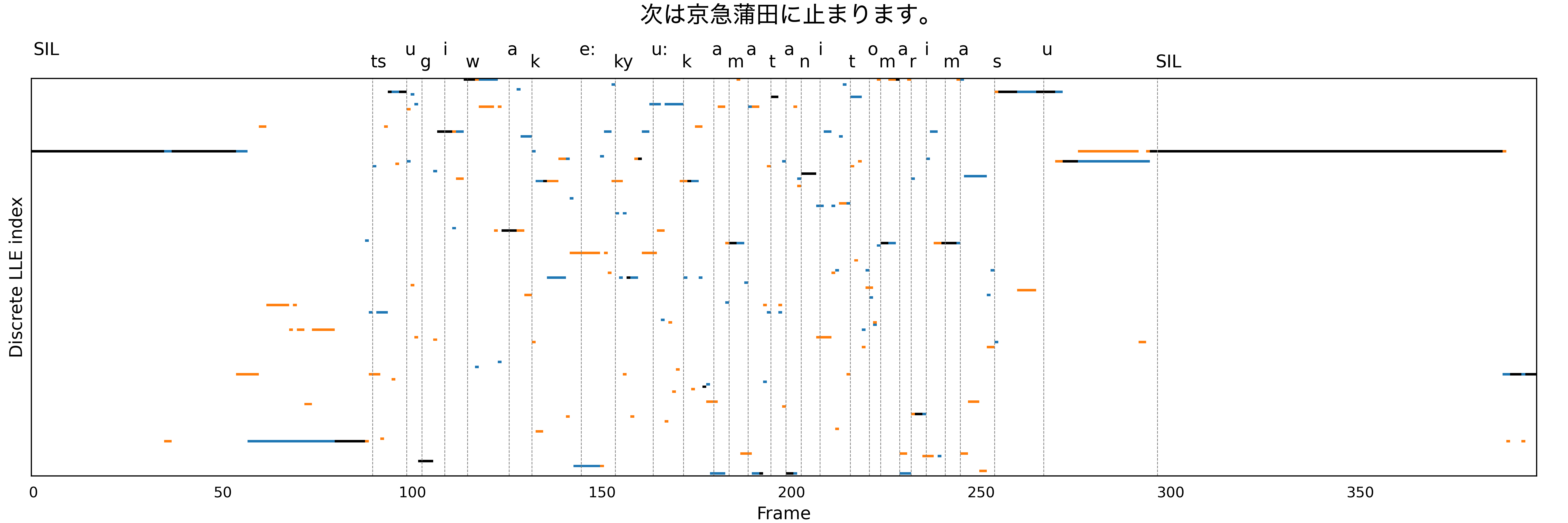}
        \vspace{-5mm}
        \caption{Examples of the 160-code discrete LLE sequences. One utterance of a source speaker was used to generate the speech-encoded LLE (orange), while text (phoneme) and alignment information extracted from the same utterance was used to generate the text-encoded LLE (blue). The color black indicates the overlap of the speech-encoded and text-encoded LLE sequences, which covers 54.41\% of this particular example utterance.}
        \label{fig:codemap}
\vspace{-4mm}
\end{figure*}

\subsection{Speech data}

Previously, we conducted experiments with English as the target language \cite{luong2020nautilus}. In this work, we used Japanese to test our methodology under a new condition. Specifically, several native female Japanese speakers, as listed in Table \ref{tab:data}, were selected as the target speakers.
The Japanese model was first initialized on a large-scale low-quality transcribed speech corpus with a diverse linguistic content. We used $\sim$236 hours of speech (978 speakers) from the 16 khz Corpus of Spontaneous Japanese (CSJ) \cite{maekawa2000spontaneous} for this purpose. Then, we fine-tuned it on a quality-controlled transcribed speech corpus for the desired sampling rate. We used $\sim$134 hours of speech (235 speakers) from an in-house 24 khz Japanese Voice Bank Corpus for this step.
The same policy was applied for the training of both NAUTILUS and NAUTILUS-VQ, and it is similar to the one we used when we trained the English models \cite{luong2020nautilus}.
The well-trained modules were used to adapt to the target speakers using their untranscribed speech data. For the first scenario, five speakers with about 45 minutes of speech were used to compare the performances of NAUTILUS and NAUTILUS-VQ.
For the second scenario, a different speaker, XL10, was used to investigate the performances when adapting with different amounts of data. This speaker was chosen because she was included in our previous experiments using the conventional TTS system \cite{luong2019training}.

\section{Evaluations}
\label{sec:evaluations}

\subsection{Vector quantization latent spaces for voice cloning}

We first compare the original and the VQ-based systems on the voice cloning task: specifically their perceptive evaluations for TTS and VC\footnote{Samples are available at \url{https://nii-yamagishilab.github.io/sample-preliminary-nautilus-vq/}}. We used about 45 minutes of untranscribed speech of the first five target speakers in Table \ref{tab:data} for this scenario. This amount of data was more than our previous English experiments \cite{luong2020nautilus}, which were conducted with just five or ten minutes of speech.
We used a crowdsourcing service to conduct the survey. A total of 241 native speakers, each of whom did one to five sessions, participated. Listeners were asked to pick the preferred sample between two presented in terms of either quality or speaker similarity (which includes a reference sample). The results are shown in Fig.\ \ref{fig:sub1}.  In total, each speaker/system/task was judged 300 times.
Interestingly, the NAUTILUS-VQ system had worse results for most speakers except F001. As most of the differences between the two systems were marginal, we conclude that the proposed vector quantization latent space can be used for voice cloning but has a small degradation in quality.

\subsection{TTS speaker adaptation with different amount of untranscribed speech data}

Next, we investigate the performances of both NAUTILUS and NAUTILUS-VQ for TTS unsupervised speaker adaptation with varying amount of adaptation data.
For this scenario, we use speaker XL10, with whom we have more than 13 hours of speech data.  Previously, we found that with this amount of data, the SD system of this speaker does not seem to benefit from the joint training with augmented data from other speakers \cite{luong2019training}.
In addition to the natural utterances (NAT) and generated utterances from our voice cloning systems, we included generated utterance from the conventional TTS system trained on 12.9 hours of transcribed speech of XL10. Basically, we reused the SD system in \cite{luong2019training} as the upper bound for this experiment.
The same listeners who participated in the first scenario survey were asked to do this one as well. Specifically, they were asked to judge the naturalness of a speech sample in a typical 5-point scale mean opinion score (MOS) question, and the likelihood that two presented samples were spoken by the same person on a 4-point scale. 
The results are shown in Fig. \ref{fig:sub2}. As expected, none of the TTS systems were as good as natural speech, but the speaker similarity was not too far behind.
Second, none of our voice cloning systems were as good as the SD baseline, which is not surprising as SD was trained on 12.9 hours of transcribed speech \cite{luong2019training} while our systems cloned voices with only a small amount of untranscribed utterances.
Third, between our systems, NAUTILUS-VQ has worse results than NAUTILUS in most data points, but not significantly so.
Finally, the most interesting results were the performances when adapting to varying amounts of data, where we found that the performance of the NAUTILUS system seemed to peak at 125 utterances.
These findings demonstrate the potential of our voice cloning system but at the same time reveal its remaining limitations.

\subsection{Visualization of the discrete LLE sequences}

One advantage of our speech synthesis is its ability to use as TTS and VC while maintaining a relatively consistent performance when switching between the two. 
Figure \ref{fig:codemap} shows the LLE sequences generated from text input and an utterance spoken by a source speaker not included in the training and adaptation stages. 
As the discrete LLE is forced to be one of 160 jointly trained vectors, it is easier and more intuitive to evaluate the consistency between the text-encoded and speech-encoded LLE sequences compared with the continuous representation (Fig.\ 8 in \cite{luong2020nautilus}).
Figure \ref{fig:codemap} shows the discrete LLE sequences generated by the text and speech encoders using text and speech of the same content. For a perfectly consistent TTS/VC system, we expect the two sequences to be perfectly matched but it was not the case as seen in Fig.\ \ref{fig:codemap}. Most of the overlap occurred at the start and the end of the utterance, which is the silence phoneme, but not much elsewhere.
Overall, the LLE sequences were sparse and fragmented. For further improvement, focusing on condensing the latent space and stabilizing the text-encoded and speech-encoded LLE sequences would be a good direction. 

\section{Conclusion}
\label{sec:conclusion}

In this paper, we investigated the feasibility of using VQVAE-based components to train a discrete latent linguistic embedding for a consistent performance TTS/VC system.
While the perceptive evaluations showed that the proposed NAUTILUS-VQ system is not as good as the original system, having these different approaches to model the linguistic latent spaces is handy for many practical reasons.
Understanding the dynamics of different methods is also important for the development of a more sophisticated speech synthesis system that can solve more complex and elaborate tasks, such as controlling speaking style \cite{wang2018style}, denoising TTS \cite{zhang2021denoispeech}, or generating audio other than speech \cite{engel2019gansynth}.
As VQVAE is just one way to model a jointly trained discrete latent space, other methods \cite{jang2016categorical,louizos2018learning} or assumptions \cite{ding2019group,ho2020non} about the nature of the latent space may lead to different results and have different utilities for specific application scenarios.

\section{Acknowledgements}

This work was partially supported by a JST CREST Grant (JPMJCR18A6, VoicePersonae project), Japan, MEXT KAKENHI Grants (18H04112, 21H04906), Japan, and KDDI research, Japan.

\bibliographystyle{IEEEtran}

\bibliography{main}

\end{document}